Monte Carlo algorithm for calculating pre-neutron fragment mass and kinetic energy distributions from measurements using the 2E Technique for reaction $^{235}$U(n$_{th}$, f).

M. Montoya, M. Roca and M. Álvarez

Universidad Nacional de Ingeniería, Av. Túpac Amaru 210, Rímac, Lima, Peru

Abstract

A nuclear fission event of an actinide results in the formation of two complementary primary fragments with masses $(A, A')$, which subsequently acquire kinetic energies $(E, E')$ due to Coulomb repulsion. Following this, they emit $(n, n')$ prompt neutrons, isotropically relative to their respective centers of mass, each with their respective kinetic energies $(\eta_1, \eta_2, \ldots \eta_n; \eta'_1, \eta'_2, \ldots, \eta'_{n'})$. Consequently, due to recoil effects, the fragments reach the detectors with altered kinetic energies $(e, e')$. This study simulates an experiment using the $2E$ technique, intending to measure the distribution of $(e, e')$ and $(n, n')$, from which it aims to infer the distribution of $A, E$ and the average prompt neutron multiplicity as a function of pre-neutron fragment mass $\bar{\nu}(A)$. For this purpose, a distribution of primary quantities is assumed as input to a Monte Carlo simulation algorithm of the experiment, whose output data should reproduce the values observed in that experiment.

**Introduction**

In neutron-induced fission studies, especially with thermal neutrons, the crucial quantities to measure relative to the pre-neutron (primary) fragment mass $(A)$ are the mass yield $Y(A)$, the average total kinetic energy $(\overline{TKE}(A))$, where $TKE = E + E'$, its corresponding standard deviation $\sigma_{TKE}(A)$, and the average prompt neutron multiplicity $\bar{\nu}(A)$ [1][2][3][4][5][4][6][7] [8] [9] [10]. These measurements are fundamental for understanding nuclear fission dynamics and are crucial for practical applications in nuclear energy, nuclear safety, and nuclear physics research.

In 2019, using the Monte Carlo simulation method for the reactions $^{233}$U(n$_{th}$, f) and $^{235}$U(n$_{th}$, f), Montoya demonstrated that the measured curve of the average prompt neutron multiplicity $(\bar{\nu})$ is highly dependent on the technique used [11]. In the same year, Montoya and Romero, employing a similar method, found that for the spontaneous fission of $^{252}$Cf, the $\bar{\nu}$ curve as a function of the provisional mass, is overestimated in comparison to the curve as a function of pre-neutron mass, particularly around the mass region of 122 [12]. In 2020, studying the reaction $^{239}$Pu(n$_{th}$, f), Montoya showed that the $\bar{\nu}$ curve simulated as measured by the 1V1E technique is overestimated compared to the curve assumed as a function of the primary mass in the mass region around 116. [1].

In 2020, Al-Adili et al. demonstrated the absence of an accurate correlation between fragment data and neutron data. They proposed a new evaluation of the prompt neutron multiplicity as a function of mass for the reaction $^{235}$U(n$_{th}$, f) [10].

In this work, we propose a Monte Carlo simulation algorithm to evaluate the relationship between the curve of average prompt neutron multiplicity $\bar{\nu}$ as a function of primary mass, and that as measured by the 2E technique, that is expected to be the same.

## Methodology

In a fission event of a nucleus with mass $A_f$, the pre-neutron masses of the complementary fragments $(A, A')$ are defined at the scission point, and they obey the relationship:

$$A_f = A + A'. \tag{1}$$

After Coulomb repulsion, the fragments acquire kinetic energies $(E, E')$, which obey the momentum conservation relationship [13]:

$$AE = A'E'. \tag{2}$$

A first approximation for the final kinetic energy of complementary fragments, is obtained neglecting the recoil effect due to neutron emission [1] [11]. Under that condition, the final values of the kinetic energy of the complementary fragments would be given by the following relations:

$$e = E\left(1 - \frac{n}{A}\right), \tag{3a}$$

and

$$e' = E'\left(1 - \frac{n'}{A'}\right). \tag{3b}$$

Here $(n, n')$ are the numbers of prompt neutrons emitted from the respective fragments.

In this work, we examine the case of thermal neutron-induced fission of $^{235}$U by simulating the recoil effect due to neutron emission to approximate the pre-neutron kinetic energy of the complementary fragments. We simulate an isotropic neutron emission relative to the center of mass of the emitter fragment. Utilizing values of $(e, e')$, which differ from those provided by relations 3a and 3b, we calculate the corresponding provisional masses $(A^*, A^{*\prime})$ using the mass and momentum conservation relations:

$$A_f = A^* + A^{*\prime} \tag{4a}$$

and

$$A^* e = A^{*\prime} e'. \tag{4b}$$

In our approach, we employ a Monte Carlo simulation algorithm. The input data for this algorithm include the yield $Y(A)$, the average total kinetic energy ($\overline{TKE}(A)$) and the average prompt neutron kinetic energy $\bar{\eta}(A)$ which are sourced from the experimental findings of Al-Adili et al. [10]. Furthermore, we base the values for the derivative of the average total kinetic energy with respect to the average neutron multiplicity $\partial \overline{TKE}/\partial \bar{\nu}$ on the data provided by Nishio et al. [6]. Additionally, the curve representing the standard deviation of the total fragment kinetic energy distribution, denoted as $\sigma_{TKE}(A)$ is derived from the results presented in Ref. [14].

In each simulated fission event, when we have a fragment with mass $A$ and energy $TKE$, the number of prompt neutrons emitted by the complementary fragments is determined by rounding down the following values:

$$n(A, TKE) = \bar{\nu}(A)\left(1 + \frac{TKE - \overline{TKE}}{\alpha(A)} + \frac{r}{3}\right) + 0.5 \tag{5a}$$

and

$$n'(A', TKE) = \bar{\nu}(A')\left(1 + \frac{TKE - \overline{TKE}}{\alpha(A')} - \frac{r}{3}\right) + 0.5. \tag{5b}$$

Here, $r$ follows a Gaussian distribution with a mean of 0 and a standard deviation of 1. This parameter, which has opposite signs for complementary fragments, indicates an anti-correlation between the multiplicities of prompt neutrons from these fragments.

We assume that the fragments emit isotropically $(n, n')$ neutrons, each with kinetic energies denoted as $(\eta_1, \eta_2, \ldots \eta_n; \eta'_1, \eta'_2, \ldots \eta'_{n'})$. All these neutrons possess the average kinetic energy $\bar{\eta}(A)$ relative to the emitter fragment center of mass [10]. Additionally, every time a neutron is emitted, the emitting fragment undergoes a recoil, which affects both its kinetic energy and its direction of motion. Consequently, the final complementary fragments that are detected post-neutron emission have altered kinetic energies, represented by $(e, e')$.

Using the values of $(e, e')$, we calculate the provisional masses $(A^*, A^{*'})$, applying the conservation of mass and momentum principles as outlined in relations 4a and 4b. It's important to note that this calculation is an approximation, as it disregards the exact impact of recoil resulting from the emission of prompt neutrons.

In our simulation, we modeled the average prompt neutron multiplicity curve, $\bar{n}(A)$, without incorporating any peak at the primary mass around $A = 112$. A significant discrepancy between the curves representing the average prompt neutron multiplicity as a function of provisional mass $\bar{n}(A^*)$ and as a function of pre-neutron mass $(\bar{n}(A))$, respectively, is found as result of the simulation. The values of $\bar{n}(A^*)$ align closely with the experimental findings of Göök et al. [9]. Notably, these values display a peak at mass 112, which is 1 unit higher than the corresponding $\bar{n}(A)$ value. See Fig. 1.

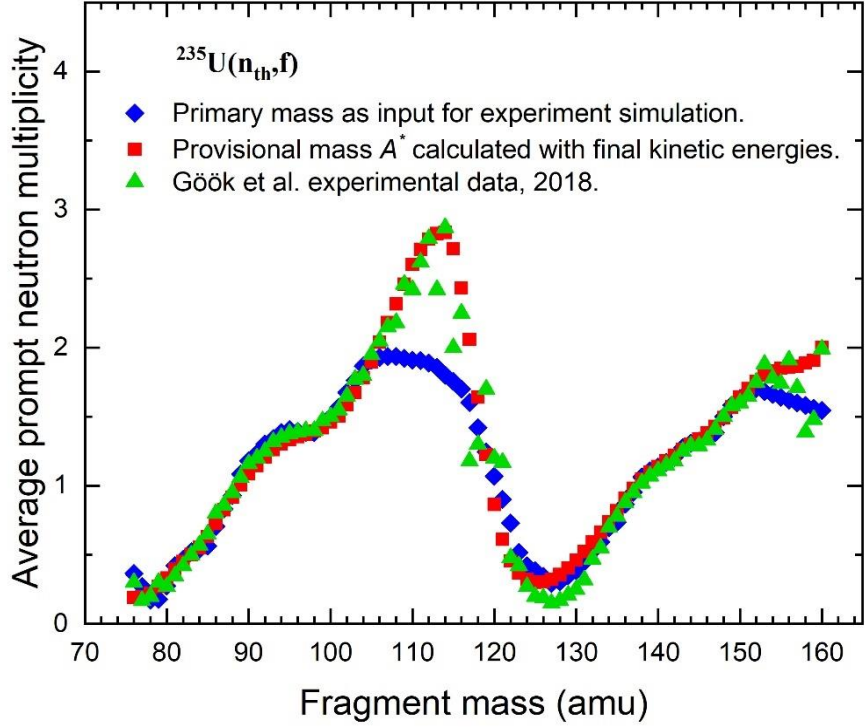

Figure 1: Simulated average of the average prompt neutron multiplicity as a function of the primary fragment mass (diamonds), as a function of provisional mass (squares), and the results of the experiment by Göök et al. [9]. (triangles).

Based on relations 3a and 3b, we modify our approach by using the average prompt neutron multiplicity instead of the number of emitted neutrons. This is done as a function of provisional mass and final total kinetic energy. Using this method, we calculate the approximate pre-neutron values of the kinetic energies of the complementary fragments through the following relations:

$$E_{\bar{n}} = \frac{e}{1 - \bar{n}(A^*, tke)/A^*} \tag{6a}$$

and

$$E'_{\bar{n}} = \frac{e'}{1 - \bar{n}'(A^{*\prime}, tke)/A^{*\prime}}. \tag{6b}$$

These equations allow us to estimate the kinetic energies of the fragments before neutron emission, considering the average neutron multiplicity relative to their respective provisional masses and final total kinetic energies.

Following this methodology, we then compute the approximate pre-neutron masses of the complementary fragments using these relations:

$$A_{\bar{n}} = A_f \frac{E'_{\bar{n}}}{E_{\bar{n}} + E'_{\bar{n}}} \tag{7b}$$

and

$$A'_{\bar{n}} = A_f \frac{E_{\bar{n}}}{E_{\bar{n}} + E'_{\bar{n}}}. \tag{7b}$$

The values $E_{\bar{n}}$ and $E'_{\bar{n}}$ are approximate estimates. Consequently, the relations 7a and 7b, which rely on these values, are also approximate in nature. By employing relations 6a, 6b, 7a, and 7b, we are able to calculate the primary masses $A_{\bar{n}}$ from provisional masses $A^*$. The resulting curves for $\bar{n}(A_{\bar{n}})$ are illustrated in Figure 2. Notably, the peak of the curve $\bar{n}(A_{\bar{n}})$ in the region around $A = 112$ is observed to be 0.5 units lower than the corresponding peak in the curve $\bar{n}(A^*)$.

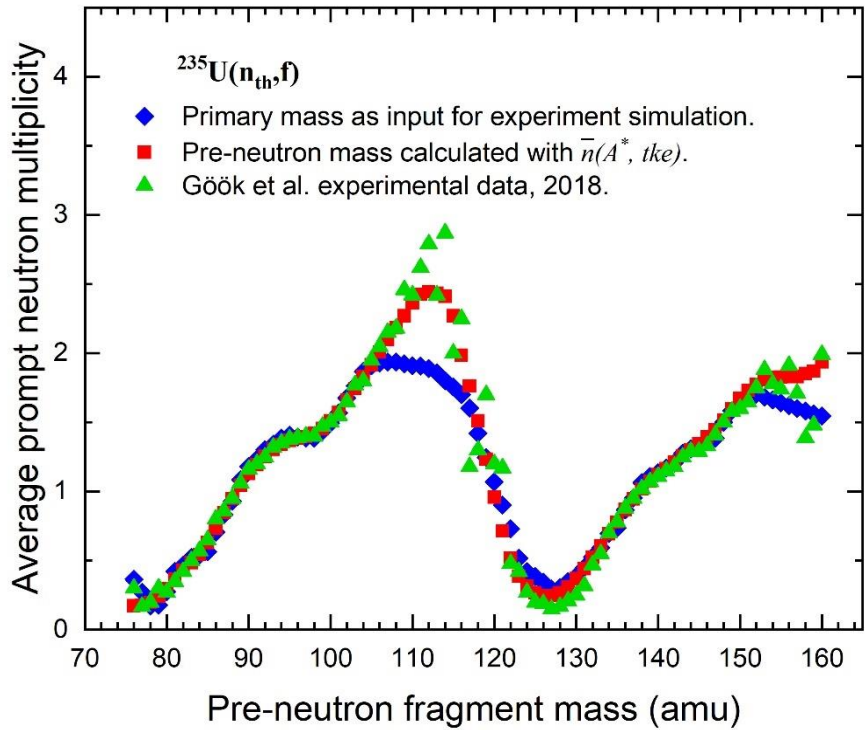

Figure 2: Simulated average of the average prompt neutron multiplicity as a function of the mass of primary fragments (diamonds), results from the simulated experiment using

the $2E$ technique taking into account the average prompt neutron multiplicity as a function of fragment provisional mass and kinetic energy to calculate pre-neutron mass (squares), and the results of the experiment by Göök et al. [9]. (triangles).

In relations 6a and 6b, we make a modification by substituting the average neutron multiplicities, $\bar{n}(A^*, tke)$, with the integer count of neutrons emitted, $n(A^*, tke)$. This adjustment results in the derivation of mass values denoted as $A_n$. When comparing the resulting curve $\bar{n}(A_n)$, with the original curve $\bar{n}(A)$, we find that $\bar{n}(A_n)$ is closer in alignment than the curve $\bar{n}(A_{\bar{n}})$. We observe a proximity of the curve $\bar{n}(A_n)$ to the experimental results obtained by Al-Adili et al. [10]. See Fig. 3.

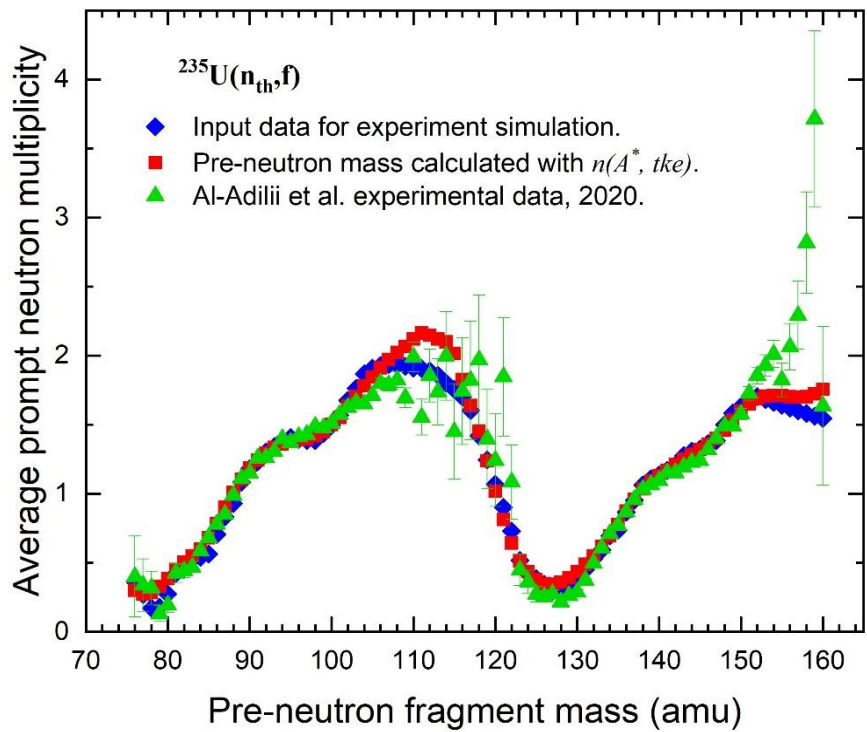

Figure 3: Simulated average of the average prompt neutron multiplicity as a function of the mass of primary fragments (diamonds), results from the simulated experiment using the $2E$ technique taking into account the integer number of emitted neutrons to calculate the pre-neutron mass (squares), and the results of the experiment by Göök et al. [9]. (triangles).

**Discussion**

The primary source of inaccuracy in the simulated $2E$ technique stems from the fact that relations 6a and 6b, which are used to calculate the pre-neutron kinetic energy (denoted exactly equal the actual mass $A$. This discrepancy between $A_{\bar{n}}$ and $A$ causes a dispersion in the calculated average prompt neutron multiplicity $\bar{n}(A_{\bar{n}})$ as compared to the average neutron multiplicity based on the primary mass of the fragments, $\bar{n}(A)$. This means that the differences in kinetic energy estimates influence the accuracy of neutron multiplicity as a function of fragment mass calculations in the simulation.

Our algorithm does not consider resolution in the measurement of kinetic energy. This resolution is crucial for precise calculations. If these experimental kinetic energy resolutions had been included, the resulting curves would have deviated even more from the curves based on primary mass.

Furthermore, $\bar{n}(A_{\bar{n}})$ is a multivariable function that depends on the mass distribution, the kinetic energy of the fragments, the number of emitted neutrons, and the kinetic energies of each neutron, which measured values lack or accuracy.

In summary, to accurately reconstruct the $\bar{n}(A)$ curve, it is necessary to simulate the distribution of all quantities related to the fragments. These simulated inputs for the algorithm should be such that the resulting average prompt neutron multiplicity aligns with the experimental values obtained from the 2E technique or other similar techniques.

During the preparation of this work the authors used ChatGPT to enhance the composition of the original sentences. After using this tool, the authors reviewed and edited the content as needed and takes full responsibility for the content of the publication.